\def\nref  {\noindent\parshape 2 0.0 truein 06.5 truein 0.4 truein 06.1 truein}
\def\AA{\leavevmode\setbox0=\hbox{h}\dimen1=\ht0 \advance\dimen1 by-1ex
 \rlap{\raise.67\dimen1\hbox{\char'27}}A }                                 

\def\lsim{$_{\sim}^{<}$}

\def\\{$\backslash$}
\def\hh       {{$^{h}$}}
\def\mm       {{$^{m}$}}
\def\ss       {{$^{s}$}}                  
\def\deg      {{\ifmmode^\circ\else$^\circ$\fi} } 
\def\arcm     {{\ifmmode {'\ }\else$'     $\fi} } 
\def\arcs     {{\ifmmode{''\ }\else$''    $\fi} } 

\def\cle      {{$_ <\atop{^\sim}$}}

\def\nref  {\noindent\parshape 2 0.0 truein 06.5 truein 0.4 truein 06.1 truein}

\def\nref  {\noindent\parshape 2 0.0 truein 06.5 truein 0.4 truein 06.1 truein}

\documentstyle [12pt,aasms4]{article}      

\begin{document}

\title{\bf Optically Faint Microjansky Radio Sources}

\author{E. A. Richards\footnotemark[1]}
\footnotetext[1]{present address: Arizona State University, Department of 
Physics \& Astronomy,
Tempe, AZ 85287-1504}

\affil{National Radio Astronomy Observatory\footnotemark[2] and University of Virginia}
\footnotetext[2]{The National Radio Astronomy Observatory is a facility
of the National
Science Foundation operated under cooperative agreement by Associated
Universities, Inc.}

\author{E. B. Fomalont \& K. I. Kellermann}
\affil{National Radio Astronomy Observatory, 520 Edgemont Road, Charlottesville, 
Virginia 22903 \newline
electronic mail: kkellerm@nrao.edu, efomalon@nrao.edu}

\author{R. A. Windhorst}
\affil{Arizona State University, Departemnt of Physics \& Astronomy,
Tempe, AZ 85287-1504 \newline
electronic mail: rogier.windhorst@asu.edu}

\author{R. B. Partridge}
\affil{Haverford College, Department of Astronomy, 
 Haverford, PA 19041 \newline
electronic mail: bpartrid@haverford.edu}

\author{L. L. Cowie \& A. J. Barger}
\affil{University of Hawaii, 2680 Woodlawn Dr.,Honolulu, HI 96822 \newline
electronic mail: cowie@ifa.hawaii.edu , barger@ifa.hawaii.edu}

\slugcomment{Accepted for Publication in the Astrophysical Journal
  Letters}

\begin{abstract}

    We report on the identifications of radio sources from our
survey of the Hubble Deep Field and the SSA13 fields, both of
which comprise the deepest radio surveys to date at 1.4 GHz and
8.5 GHz respectively.
About 80\% of the
microjansky radio sources are associated with moderate redshift
starburst galaxies or AGNs within the I magnitude range of 17 to 24
with a median of I = 22 mag. Thirty-one (20\%) of the radio sources
are:
1) fainter than $I>$25 mag,
with two objects in the HDF $I_{AB}>$28.5, 2) often identified with very red
objects $I-K>$4 , and 3) not significantly different in radio
properties than the brighter objects.  We suggest that most of these
objects are associated with heavily obscured starburst galaxies with
redshifts between 1 and 3.  However, other mechanisms are discussed and
cannot be ruled out with the present observations.

\end{abstract}


\slugcomment{Submitted to {\it The Astrophysical Journal Letters}, November
1998}

\section{Introduction}

	We have recently completed a comprehensive radio
survey of the Hubble Deep Field (HDF) region using
the Very Large Array. 
Within a 40\arcmin ~region centered on the HDF we detected 40 sources at 8 
GHz
above 8 $\mu$Jy and 371 sources at 1.4 GHz above 40 $\mu$Jy
(Richards et al. 1998, Richards 1999a).
The angular resolution of the VLA observations ranged from 2-6\arcsec ~
and subsequent 1.4 GHz imaging with the Multi-Element 
Radio Linked Interferometer (MERLIN) increased the 
resolution to 0.2" (Muxlow et al. 1999; Richards 1999b). 

	Another deep radio survey conducted by our group (Windhorst et 
al., 1995; Kellermann et al. 1999) was centered on the
Small Selected Area 13 field ($\alpha$ =13\hh 12\mm 17.4\ss and
$\delta$ = +42\deg 38\arcmin \hskip5pt 05\arcsec, J2000) and employed the VLA C and
D configurations at 8.4 GHz producing a 6\arcsec ~beam.
The rms noise in this image is 1.5 $\mu$Jy with
39 radio source catalogued as part of a complete 5$\sigma$ sample.

	At the 10-100 $\mu$Jy flux level, 60\%
of the radio sources are identified with bright (I $\sim$ 22 mag)
disk galaxies (irregulars, peculiars, mergers),
with typical redshifts of 0.2-1 (Richards et al. 1998). 
The diffuse, steep spectrum emission
observed in these galaxy systems is interpreted
as optically thin synchrotron radiation from coalesced
supernovae remnants with perhaps a smaller fraction of
bremsstrahlung radiation from HII regions. Some 20\% of the 
microjansky radio sample are associated with
low luminosity AGN (weak FR Is, Seyferts, and ellipticals).

	The remaining 20\% of the microjansky radio sources
cannot be identified to optical magnitude limits of
I = 25 mag. Three of these radio sources are contained in the
Hubble Deep Field of which two remain unidentified at $I_{AB} >$ 28.5 
mag. In this 
Letter we list the sample of these radio sources associated
with faint and/or red objects and discuss their possible
nature.

\section{Identification and Colors of Radio Sources}

	Optical photometry in the I and K bands were
obtained in the HDF (Barger et al. 1999) and SSA13
fields (Cowie et al. 1996). First we aligned
these wide field I and K band images under our
radio detections with 
residual astrometric uncertainties
of approximately 0.2\arcsec ~. The absolute radio positions are
known to 0.1-0.2\arcsec ~rms in the HDF and
0.3-0.6\arcsec ~rms in SSA13 and are tied to the standard
FK5 inertial frame (Richards et al. 1998).

        Using our astrometrically aligned optical images, we
attempted to make identifications of the 111 (79 in the HDF, 32 in
SSA13) radio sources
contained within the I-band images which only partially
cover the radio fields. These optical
images have comparable depths of 24.8 mag and 25.0 mag (2 $\sigma$)
in the Kron-Cousins system within the HDF and SSA13 fields,
respectively. Identifications were chosen based on optical magnitude and
the optical-radio angular separation as 
described in Richards et al. (1998).
In most cases the radio emission
lies directly on top of a bright galaxy and the
association is unambiguous. 
For any identification reliability less than
10\% , we classified as 'unidentified,'.
The Hubble Space Telescope flanking field images
were also checked for optical emission for
each of these unidentified sources in the HDF
region.

     A magnitude histogram of the optical identifications is 
shown in Figure 1. Eighty-four secure identifications
were found. Clearly, 
the bulk of the sample is identified with
relatively bright (I \cle  22 mag) galaxies.
As can be seen from Figure 1,
approximately 20\% of the radio sources
cannot be identified to I = 25 mag, although 
a few of the unidentified radio sources
show faint optical emission below the formal
completeness limit of the optical image.

	We next used deep HK' images of the
HDF (Barger et al. 1999) and SSA13 (Cowie et al. 1996)
to estimate K magnitudes (K = HK' - 0.3 mag in the Johnson
system) 
for our radio detections. I--K colors for the 84 optical 
identifications (heavy dots)   
are compared against the I--K colors of the 
field population (light dots; Barger et al. 1999) in Figure 2.

	Two properties of the identifications of the
84 radio sources are clear from Figure 2: 1) For
objects fainter than I $>$ 21 mag, the galaxies associated
with the identified radio sources are redder than the general
field galaxy population at the same flux limit. 2) There is a monotonic
trend for the optically fainter
radio sources to have redder colors and reaches $I-K \simeq$
3 at I = 24 mag.

\section{The Sample of Optically Faint and Red Radio Sources}

        Using the list of 111 radio objects with
existing optical/near-infrared photometry
, we isolated a sample of
optically faint and/or red radio objects with
I $>$ 25 and/or $I-K >$ 4 (two magnitudes redder 
than the mean of the field galaxies in the
HDF region; Barger et al. 1999).
Table 1 presents the 31 optically faint/red
objects along with radio flux density (measured at 1.4 GHz
for the HDF sample, and 8.4 GHz for the SSA13 sample),
statistical significance
of the radio detection, radio spectral index for the HDF
sample, and $I-K$ color.
We show radio-optical overlays of the 17
radio sources contained in the HDF region in Figure 3.
An additional red radio object in SSA13 is reported
by Partridge, Eppley, and Spillar (1999).


	Here we note distinctive properties of
a few of the optically faint radio properties,
including possible associations at other 
wavelengths. Another very red radio source listed
in Table 1 (VLA J123642+621331) is described in a companion
paper by Waddington et al. (1999).

{\bf VLA J123646+621226:} Contained in the HDF (Williams et al. 1996) and the
  deep near-infrared observations of Thompson et al. (1999),
  places magnitude limits of $I_{AB} >$28.5, $J_{110}>$29, and $H_{160}>$29 mag
  for any counterpart. The radio emission is particularly diffuse
  and has a steep radio spectrum with $\alpha$ = 0.90.
  VLA positions measured independently at 1.4 GHz and 8.5 GHz
  agree with 0.4\arcsec , consistent with their expected
  standard errors.

{\bf VLA J123651+621221:} The higher resolution HDF 
   frame clearly shows that this radio source is
   not associated with the foreground spiral to the
   south (Richards 1999c, fig. 1). Detected at $I_{AB}
   = 27.8$ mag in the HDF, this radio source is the
   second reddest optically selected object in the
   HDF-NICMOS survey of Dickinson et al. (1999)
   with $J_{110} - H_{160}= 1.6$.
   The sub-mm source HDF850.1 lies 5\arcsec ~to the northeast
   (Hughes et al. 1998; Downes et al. 1999), which has
   been detected at 1.3 mm with subarcsec
   positional accuracy. The 1.3 mm source is coincident
   with the faint 8.5 GHz radio detection VLA J123651+621226
   (Richards et al. 1998). However, the 1.3 mm image also shows
   a 3 $\sigma$ detection within 1\arcs of the position of
   VLA J123651+621221, suggesting the sub-mm source 
   HDF850.1 is a blend of two independent sources
   located in the 15\arcs JCMT/SCUBA beam.

{\bf VLA J123656+621207:} This very steep spectrum radio source
  ($\alpha > 1.3$) lies within
  the HDF, placing a limit of $I_{AB} >$ 28.5 mag on the counterpart.
  Deep near-infrared observations show no sign of emission
  at the position of the radio source to $J_{110} >$ 29,
  $H_{160} >$ 25, and K $>$ 24 mag (Barger et al. 1999,
  Dickinson private communication 1999).   
  The sub-mm source HDF850.2 lies 3\arcsec ~to the southwest
  and is likely the same object as
   the radio source (Hughes et al. 1998; Barger, Cowie \&
  Richards 1999).     
  


{\bf VLA J123721+621130:} A $I-K >$ 5.2 object is identified
  with this inverted spectrum radio source.
  The radio emission is likely synchrotron self-absorbed 
  from an optically obscured AGN. The optical galaxies
  to either side of the radio emission are probably unrelated.

\section{Discussion and Conclusions}

	The radio properties of these 31 optically peculiar
sources are not in themselves unusal. In general the radio sources
are extended over a few arcseconds, have steep radio spectral indices, and
are distributed over a range of flux density.
Only the faintness of their optical counterparts
sets them apart from the bulk of the microjansky radio
population. We also note that the sources reported here
are not variable on the time scales of days to months, over
which our observations extended, ruling out association
with variable AGN, gamma ray bursts, and supernovae.

	We consider several possibilities for the nature
of these radio sources in order of their likelihood:

1. High $z$-starbursts (1 $< z <$ 3): Given
   that the majority of submillijansky radio
   sources are associated with star-forming
   galaxies, it is plausible that
   a tail of the parent galaxy population
    is so obscured by dust that only
   the radio emission is visible. In
   this case we would expect the radio
   emission to be steep spectrum and
   possibly resolved. The host systems
   would likely be disk galaxies, possibly
   merging or perturbed in analogy to
   the local ultraluminous IRAS galaxies.
   Beyond redshifts
   of 2-3, our sensitivity to star-forming galaxies
   cuts off due to our flux density limits.
   VLA J123651+621221 is a candidate dust
   enshrouded system.

2. Extreme redshift AGN ($z >6$) - Another possibility, is
   that some of the radio sources with very faint
   optical fluxes are at extreme
   redshifts, where the Lyman 912\AA\ break blanks
   out the I-band continuum, placing
   them at $z >$ 6. With
   our present radio sensitivity, we could
   have detected a nominal FR I galaxy
   to $z \simeq$ 10. It is
   interesting to note that the empirically
   determined $K-z$
   relation observed for brighter radio
   galaxies implies redshifts greater than five
   for several of our radio sources (Dunlop \&
   Peacock 1990) {\it if} they are FR I radio galaxies.

3. Moderate redshift ( $z <$ 2) AGN: Field elliptical
   galaxies and Seyferts often
   display evidence for a radio AGN component.
   However, typically the optical magnitudes
   of such systems are $L*$ or greater
   and thus to escape detection to I = 25, would
   place them at redshifts in excess of two.
   VLA J123721+621130 might
   be of this variety.



4. One sided radio jets: We cannot discount the
   possibility that some of the optically
   unidentified radio sources are the brightest
   jet of a nearby but displaced optical
   galaxy. In this case
   we would expect the parent galaxy to be
   a luminous elliptical containing an AGN. The radio
   emission in this case should be extended and
   steep spectrum. Radio source VLA J123646+621226
   could possiblely fit in this category. However,
   FR Is comprise less than 5\% of the submillijansky
   radio population and are usually solidly identified with
   bright galaxies at $z$\lsim  1 (Kron et al. 1985).

	We appreciate conversations with D. Downes,
D. Haarsma, I. Waddington, and M. Dickinson.

\newpage 

\section*{Figure Captions}

\figcaption{I magnitude histogram of radio sources in
the HDF and SSA13 regions. The mean of the distribution is I = 21.6 mag,
with a median of I = 22.1 mag.
}

\figcaption{I--K color magnitude diagram for the identified radio sources
 in the HDF and SSA13 regions as shown as large solid dots. The smaller
 dots are the data from the field galaxy sample of Barger et al. (1999).
 The light broken line represents the mean I-K = 2.2 of the field galaxy sample.
 The solid line is a linear fit to the I-K colors of the radio identifications.
 The heavy broken line shows color-mag limits for I = 24 and K = 20 (HFF and SSA13 regions)
 and I = 24 and K = 23 (HDF region).
}

\figcaption{Radio-optical overlays of 17 optically faint radio objects
isolated in the HDF region. The grey scale is from a log stretch of an I-band 
image of
 Barger et al. (1999) and the contours from a 1.4 GHz VLA images of Richards (1999a).
 The radio resolution is approximately 2\arcsec ~ and 
 contours are drawn at the -2, 2, 4, 8, 16, 32 and 64 $\sigma$ level
 ($\sigma$ = 7.5 $\mu$Jy). The images are oriented with north to the
 top and east to the left and are 20\arcsec ~ on a side. 
}


\begin{references}

\nref Barger, A. J. et al. 1999, AJ, 117, 102 

\nref Barger, A. J., Cowie, L. L. \& Richards, E. A. 1999, AJ,
  submitted (astro-ph/9907022) 

\nref Cowie, L. L, Songaila, A., Hu, E. \& Cohen, J. 1996,
  AJ 112, 839 

\nref Dickinson, M. et al. 1999, ApJ, in press (astro-ph/9908083)

\nref Downes et al. 1999, A\&A , 347, 809 


\nref Hammer, F., Crampton, D., Lilly, S., Le Fevre, O. \&
   Kenet, T. 1995, MNRAS, 276, 1085

\nref Hughes et al. 1998, Nature, 393, 241

\nref Kellermann et al. 1999, in preparation

\nref Kron, R. G., Koo. D. C. \& Windhorst, R. A. 1985, A\& A, 146, 38

\nref Muxlow et al. 1999, in preparation

\nref Partridge, R. B., Eppley, J.\& Spillar, E. 1999, in preparation

\nref Richards, E. A. 1999a, ApJ, to appear in Januarary 2000 (astro-ph/9908313)

\nref Richards, E. A. 1999b, in "The Evolution of Galaxies on
  Cosmological Timescales", eds. J.E. Beckman and T.J. Mahoney,
  to appear in Astrophysics \& Space Sciences (astro-ph/9904033)

\nref Richards, E. A. 1999c, ApJL, 513, 9

\nref Richards, E. A., Kellermann, K. I., Fomalont, E. B., Windhorst, R. A.
        \& Partridge, R. B. 1998, AJ, 116, 1039

\nref Shaver, P., Wall, J. V. , Kellermann, K. I., Jackson, C. A. \& 
Hawkins, M. R. S. 1996, Nature, 384, 439

\nref Thompson, R. et al. 1999, AJ, 117, 17

\nref Waddington, I., Windhorst, R., Cohen, S. \& Partridge, R. B.
   1999, ApJ, submitted

\nref  Windhorst, R. A, Fomalont, E. B., Kellermann, K. I., Partridge, R.
B.,
      Richards, E. A., Franklin, B. E., Pascarelle, S. M.
       \& Griffiths, R. E. 1995,
      Nature, 375, 471

\nref Williams, R. E. et al. 1996, AJ, 112, 1335

\nref Windhorst, R. A, Fomalont, E. B., Kellermann, K. I., Partridge, R.
B.,
      Richards, E. A., Franklin, B. E., Pascarelle, S. M.
       \& Griffiths, R. E. 1995,
      Nature, 375, 471               

\end{references}
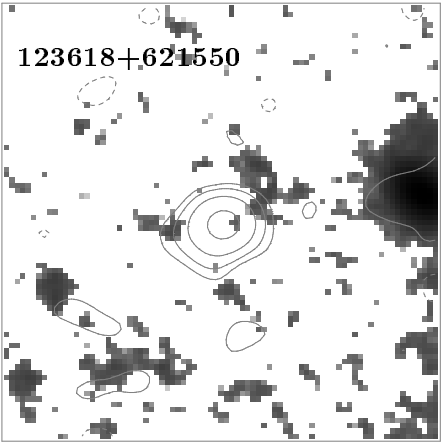
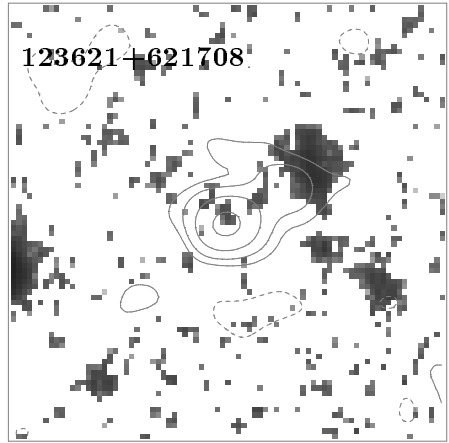
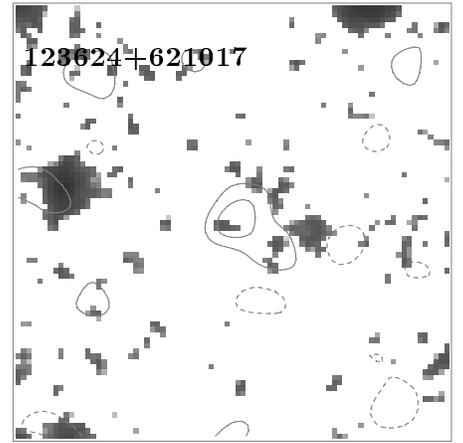
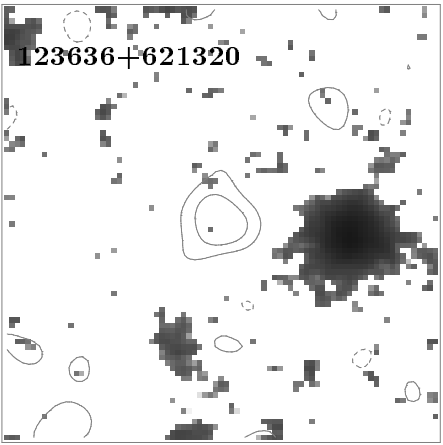
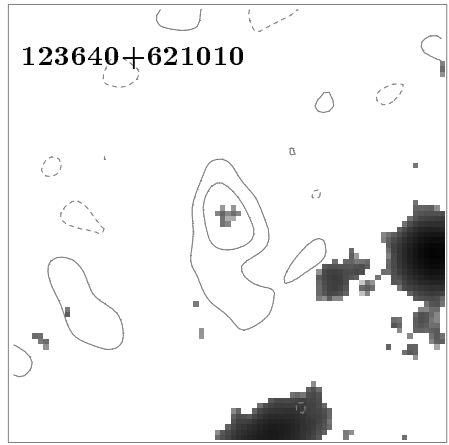
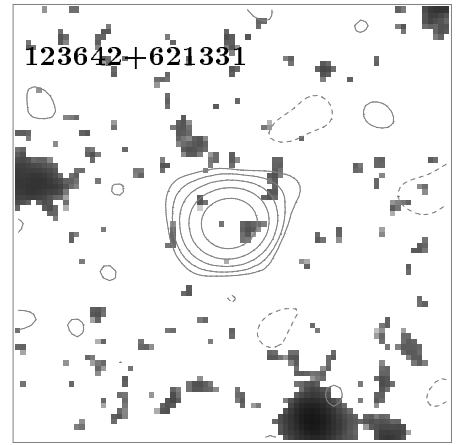
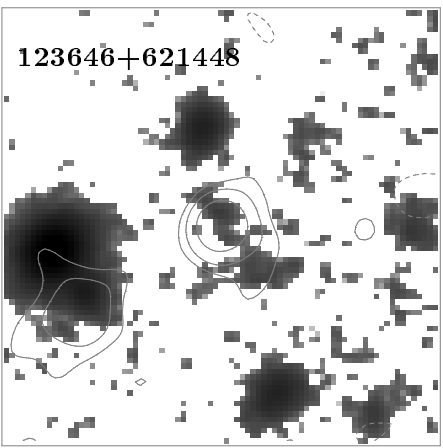
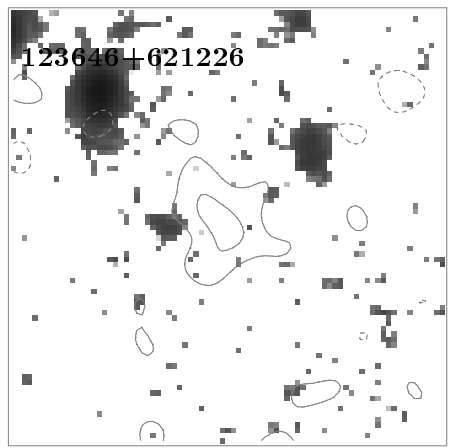
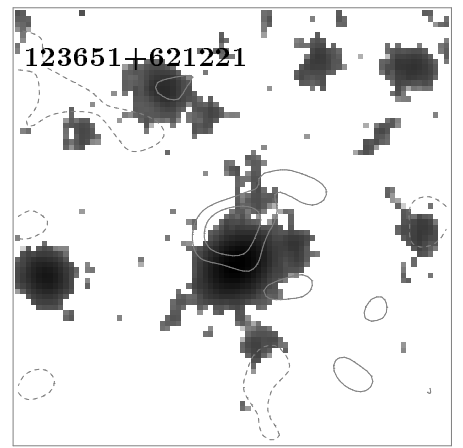
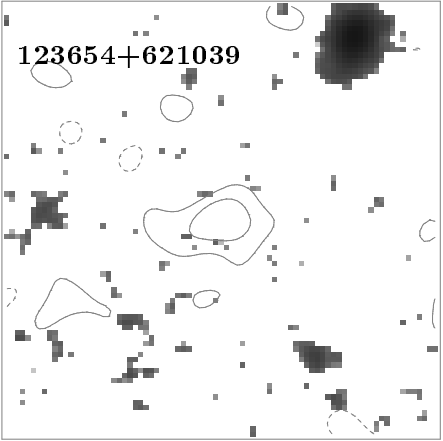
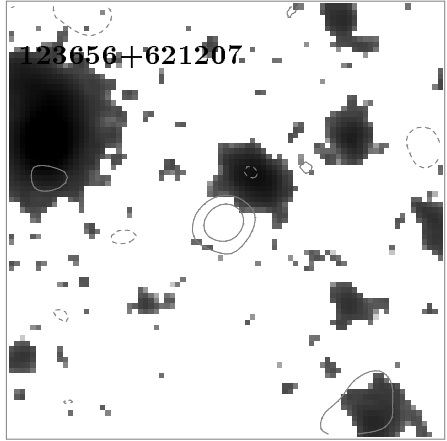
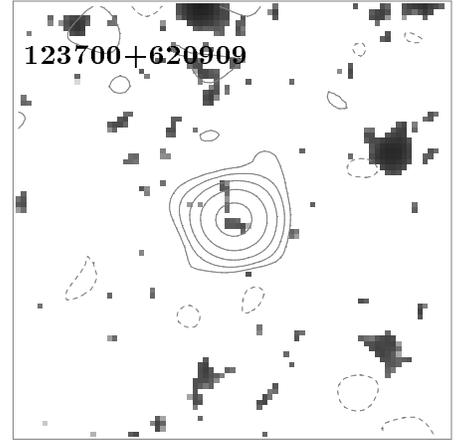
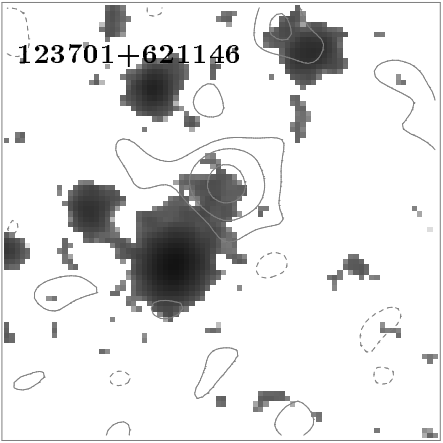
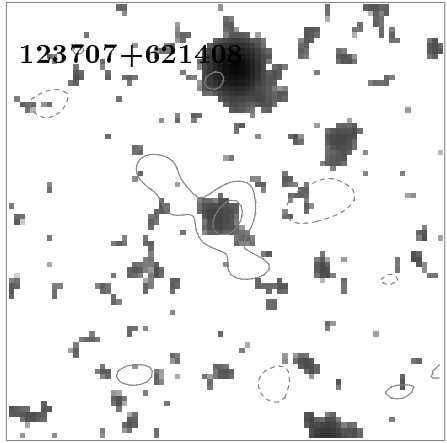
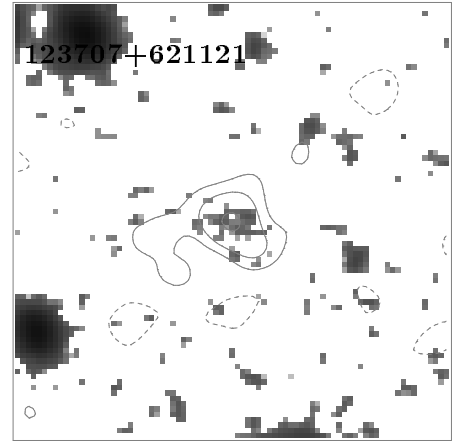
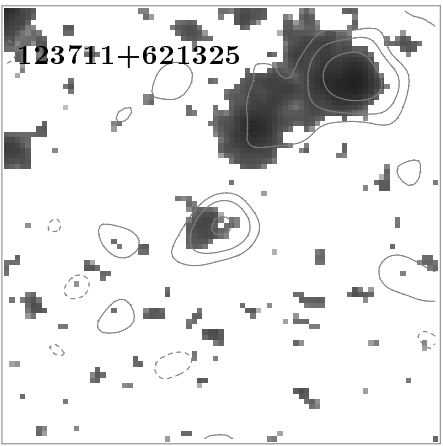
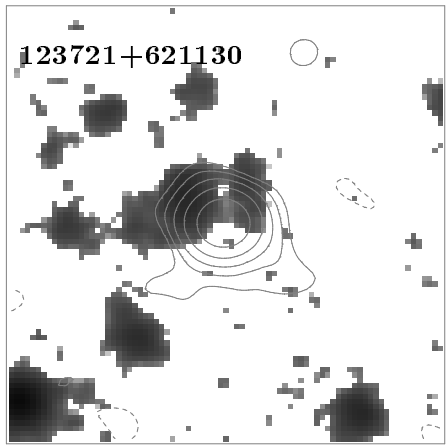
\end{document}